\documentclass[aps,amsfonts,showpacs]{revtex4}

\usepackage{amsmath}
\usepackage{graphicx}
\begin{document}

\def\be{\begin{equation}}
\def\ee{\end{equation}}
\def\bea{\begin{eqnarray}}
\def\eea{\end{eqnarray}}
\def\bml{\begin{mathletters}}
\def\eml{\end{mathletters}}
\def\l{\label}
\def\b{\bullet}
\def\eqn#1{(~\ref{eq:#1}~)}
\def\no{\nonumber}
\def\av#1{{\langle  #1 \rangle}}
\def\m{{\rm{min}}}
\def\M{{\rm{max}}}

\title{Evolutionary dynamics of the most populated genotype on rugged fitness landscapes}

\author{Kavita Jain}
\affiliation{Department of Physics of Complex Systems, Weizmann Institute of Science, Rehovot 76100, Israel}
\widetext
\date{\today}

\begin{abstract}
We consider an asexual population evolving on rugged fitness landscapes 
which are defined on the multi-dimensional genotypic space and have many 
local optima. We  
track the most populated genotype as it changes when the population 
jumps from a fitness peak to a better one during the process of adaptation. 
This is done using the dynamics of the shell model which is a simplified 
version of the quasispecies model for infinite populations and standard 
Wright-Fisher dynamics for large finite populations. We show that 
the population fraction of a genotype obtained within the 
quasispecies model and the shell model match for fit 
genotypes and at short times, but the dynamics of the two models  
are identical for questions related to the most populated genotype. 
We calculate exactly several properties of the jumps in infinite 
populations some of which 
were obtained numerically in previous works. We also present our 
preliminary simulation results for finite populations. 
In particular, we measure the jump distribution in time and find that it decays as $t^{-2}$ as in the quasispecies problem.
\end{abstract}
\maketitle

\section{Introduction}

The question of mode in evolution, especially in the context of 
speciation has engaged the attention of 
many evolutionists for over a century and continues to do so 
\cite{Fitch:1994}. The issue is whether evolution occurs by 
smooth gradual changes (gradualism) as put forwarded by Darwin, or 
 sudden large jumps 
(punctuationism) \cite{Mayr:1963, Eldredge:1972}. Some examinations of 
fossil record indicate that 
new species can arise by either mode, or even by a combination of the two 
\cite{Benton:2001}. However, a complete and unambiguous answer is hard to 
obtain at the level of macroevolution due to the incompleteness and 
irreproducibility of the fossil data. 
In the last decade or so, researchers have become interested in carrying out 
long-term evolution experiments in the laboratory 
\cite{Elena:2003a}. Typically one starts with a microbial population 
 maladapted to a given environment such as a colony of starving bacteria,  
and track its 
evolutionary trajectories for thousands of 
generations as it undergoes adaptive changes.  
The results of such experiments have been found to be consistent with 
both modes of evolution. For instance, 
large populations of RNA virus starting from a low fitness ancestor 
have been seen to gain fitness in a continuous manner \cite{Novella:1995}. 
On the other hand, the fitness of 
bacteria \emph {E. Coli} \cite{Lenski:1994,Imhof:2001} and 
RNA virus $\phi_6$ \cite{Burch:1999} show a punctuated pattern of evolution. 

\begin{figure}
\begin{center}
\includegraphics[angle=270,scale=0.4]{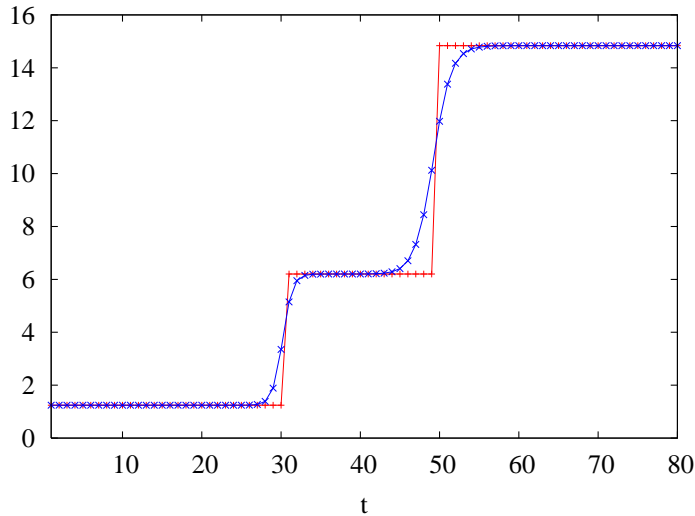}
\caption{(Color online)Punctuated change in the population fitness ($\times$) and the fitness of the most populated genotype ($+$) in a single realization of a maximally 
rugged fitness landscape. Here $L=6, N=2^{14}, \mu=10^{-6}, p(W)=W^{-2}$.}
\label{example}
\end{center}
\end{figure}

Theoretically, these results are understood using the concept of 
fitness landscape \cite{Wright:1932, Stadler:2002, Gavrilets:2004} defined 
as a map from the genotypic space into the real numbers. 
If the fitness landscape is smooth and single-peaked, starting from a low 
fitness state the population fitness increases gradually 
until it has reached the peak value \cite{Tsimring:1996}. 
The dynamics are different for a population moving 
on a rugged fitness landscape with multiple peaks \cite{Kauffman:1993}. In 
this case, the population fitness increases 
smoothly until the population 
encounters a local fitness peak where it gets trapped as a better peak 
is separated by a fitness valley. The population thus 
enters the stasis phase and waits until a favorable mutation allows it 
to shift to a better peak where it can again get localised and so on. 
Thus, the dynamics alternate between periods of 
{\it stasis} and rapid changes in fitness when the population {\it jumps} 
from a peak to an even better one. An example of this behavior is shown in 
Fig.~\ref{example}. 

That the fitness landscape 
underlying the evolutionary process is multi-peaked is supported by 
several experiments 
\cite{Korona:1994,Lenski:1994,Burch:1999,Burch:2000,Poelwijk:2007,Fernandez:2007}. 
Besides, these landscapes include biologically important and 
ubiquitous epistatic interactions \cite{Whitlock:1995} as 
the contribution of individual genes to the fitness of the genotype is 
not independent \cite{Kauffman:1993}. The class of landscapes 
considered in this work are maximally rugged and 
characterised by strong selection \cite{Jain:2007a}. 
In statistical physics, such rugged 
landscapes with a large number of local optima have appeared in the context 
of spin glass theory. Examples include random energy model \cite{Derrida:1981} 
and Sherrington-Kirkpatrick model \cite{Sherrington:1975} where the energy of 
a spin configuration plays the role of genotypic fitness and the metastability 
in the spin glass dynamics can be viewed as punctuated equilibrium \cite{Sibani:2003}.

We are interested in the statistics of jumps which occur 
when the population fitness changes rapidly. As illustrated in 
Fig.~\ref{example}, an unambiguous way of defining a jump at time $t$ 
is the change in the fitness or identity of the {\it most populated genotype} 
at $t$.  The problem of such leadership 
changes arises in several other contexts such as change in the 
position of the optimal end point of directed polymer \cite{Krug:1993}, 
highest degree node in a growing network \cite{Krapivsky:2002,Ben-Naim:2004}, 
and velocity and position of the front particle in a single-file system 
\cite{Bena:2007,Sabhapandit:2007}. 
In the following sections, before analysing the realistic case of finite 
populations, we will 
study the infinite population limit in detail. We consider the dynamics of 
Eigen's model \cite{Eigen:1971} that describes the population dynamics of 
self-replicating molecules which at low mutation rates and 
large times form a dynamic and heterogeneous quasispecies 
consisting of fittest genotype and its closely related mutants. 
The existence of such an error threshold is a generic result seen in both simple and complex fitness landscapes, and has been 
reviewed in \cite{Jain:2007b} (for recent related works, see \cite{Nilsson:2000,Kamp:2002,Saakian:2006,Park:2007}). The focus of this article is however the {\it dynamics} of the quasispecies model. 
We show that the population fraction at a genotype obtained within a 
simplified shell model \cite{Krug:2003} 
(also see \cite{Zhang:1997}) of quasispecies dynamics 
is good only for highly fit sequences and at short times. 
However, the behavior of the most populated genotype is captured correctly 
by the shell model. 
We calculate exactly some properties of the jumps (as defined above) 
within the shell model 
which were obtained numerically in previous works \cite{Krug:2003,Krug:2005,Jain:2005}. Specifically, we show that the jump distribution decays as 
$1/t^2$ in time and $1/\sqrt{k}$ as a function of the distance $k$ 
from the starting genotype for a class of fitness landscapes. 

The basic difference between a finite and infinite 
population is that 
while the former has a finite mutational spread in the genotypic space, all 
the mutants are available at all times in the quasispecies limit. 
Due to this reason, at large times, a finite population gets trapped at a 
local fitness peak and the jump probability is governed by the rate of 
stochastic tunneling \cite{Iwasa:2004,Weinreich:2005} which allows the 
population to cross the fitness valley via few low fitness mutants. 
In the quasispecies model on the other hand, the reproduction or selection 
plays the more important role than the production of mutants and 
a jump occurs when the population at a fit genotype overtakes 
the less fitter one. Although the underlying physical processes responsible for a jump are different for finite and infinite populations, we find that the jump distribution is robust in that it decays as $1/t^2$ in both the cases.

The plan of the paper is as follows. In the next section, we define a class of 
mutation-selection models. Section~\ref{quasi-shell} discusses the dynamics of 
quasispecies model and shell model. We calculate exactly the 
statistics of jumps 
within the shell model in Section~\ref{jump-shell}. These results for infinite 
population limit are compared with those obtained in simulations of finite 
population in Section~\ref{jump-finite}. Finally, we conclude with a summary 
of our results. 

\section{Models}
\l{models}

We consider a haploid, asexual population of size $N$ each of whose 
constituents carry a string 
$\sigma=\{\sigma_1,...,\sigma_L \}$ of length $L$ where $\sigma_i$ can 
assume $\ell \ge 2$ values. The string $\sigma$ can represent a genetic 
sequence ($\ell=4$), protein ($\ell=20$) or a sequence of $L$ loci with 
$\ell$ alleles \cite{Jain:2007b}. For simplicity, we will deal with 
binary sequences for which $\ell=2$ and $\sigma_i=0$ or $1$. The environment 
of the population is represented by a fitness landscape which is 
obtained by associating a non-negative real number $W(\sigma)$ to 
each sequence. 
In this article, we consider fitness $W(\sigma)$ to be a random variable 
chosen independently from a common distribution $p(W)$. This generates a 
maximally rugged fitness 
landscape with an exponentially large number (in $L$) of local maxima 
\cite{Flyvbjerg:1992,Kauffman:1993} and strong interactions amongst the loci 
\cite{Jain:2007a}.  The population evolves on this fitness landscape in 
discrete time via selection and mutation, and 
we ignore other factors responsible for genetic mixing such as recombination. 
Our model is thus applicable to microorganisms like \emph{E. Coli} and 
Hepatitis C virus which have zero or very low recombination rate 
\cite{Posada:2002}. 

Consider a population well adapted to a given environment localised around a 
peak of the fitness landscape. A change in the environment brings about a 
change in the fitness landscape, and the population will typically find itself 
in a fitness valley. We consider the adaptation process of the population 
starting from such an initial condition. 
The population fraction $X(\sigma,t)$ of sequence $\sigma$ at time $t$ 
is iterated using Wright-Fisher dynamics defined as follows. 
In each generation, an 
offspring selects a parent $p$ with probability 
$W_p(\sigma)/N \langle W \rangle$ where $W_p(\sigma)$ is the fitness of the 
parent $p$ with sequence $\sigma$ and 
$\langle W \rangle=\sum_{\sigma'} W(\sigma') X(\sigma',t)$ is the average 
fitness of the population. Then the probability $P(n)$ that parent $p$ has 
$n$ offspring in one generation is given as
\be
P(n)= {N \choose n} \left(\frac{W_p(\sigma)}{N \langle W \rangle} \right)^n 
\left(1- \frac{W_p(\sigma)}{N \langle W \rangle} \right)^{N-n}~.
\no
\ee
This implies that the average number of offspring produced equals 
$W(\sigma)/\langle W \rangle$ and hence the fitness $W(\sigma)$ has the 
physical meaning that it is proportional to the number of descendants 
produced per generation. Further, as the relative variance in the offspring 
number decays as $1/N$, it follows that the offspring number fluctuates 
from one generation to another for finite population but one can ignore these 
fluctuations arising due to random sampling for $N \to \infty$.  
After the reproduction process, point mutations are introduced independently 
at each locus of the sequence $\sigma'$ with probability $\mu$ per generation. 
Thus, a sequence $\sigma$ is obtained via mutations with the 
probability 
\be
p_{\sigma \leftarrow \sigma'} = \mu^{d(\sigma,\sigma')} (1 - \mu)^{L-d(\sigma,\sigma')}
\label{mut}
\ee
where the Hamming distance $d(\sigma,\sigma')$ is the number of point 
mutations in which the sequences $\sigma$ and $\sigma'$ differ. 

Since the selection process is not stochastic for infinite populations, 
the population frequency also does not fluctuate and 
one can work with the average frequency 
${\cal X}(\sigma,t)=\langle X(\sigma,t) \rangle$ where the averaging is 
over all realizations of the sampling process. This leads to a deterministic nonlinear equation for the fraction ${\cal X}(\sigma,t)$ \cite{Jain:2007a},
\be
{\cal X}(\sigma,t+1)= \frac{\sum_{\sigma'} p_{\sigma \leftarrow \sigma'} 
W(\sigma') 
{\cal X}(\sigma',t)}{\sum_{\sigma'} W(\sigma') {\cal X}(\sigma',t)}~
\label{quasi}
\ee
where the denominator is the normalization constant. 

\section{Dynamics of Quasispecies and shell model}
\l{quasi-shell}

The focus of this section is the quasispecies model and the related shell 
model. 
In the following, we will mainly work with the unnormalised population 
${\cal Z}(\sigma,t)$ defined as \cite{Jain:2007b}
\be
{\cal Z}(\sigma,t)={\cal X}(\sigma,t) \prod_{\tau=0}^{t-1} \sum_{\sigma'} W(\sigma') {\cal X}(\sigma',t)
\ee
in terms of which the nonlinear evolution (\ref{quasi}) reduces to the 
following linear iteration 
\be
{\cal Z}(\sigma,t+1)= \sum_{\sigma^{'}} p_{\sigma \leftarrow \sigma^{'}} 
W(\sigma') 
{\cal Z}(\sigma',t)~.
\l{linear}
\ee
Since at the start of the adaptation process the 
population finds itself at a low fitness genotype, 
we start with the initial condition 
${\cal X}(\sigma,0)={\cal Z}(\sigma,0)=\delta_{\sigma,\sigma^{(0)}}$ 
where $\sigma^{(0)}$ is a randomly chosen sequence. 
For small mutation probability $\mu$ ($\sim 10^{-3}-10^{-10}$) as seen 
in various asexual microbes \cite{Drake:1998, Jain:2007b}, after one 
iteration we have
\be
{\cal Z}(\sigma,1)=  \mu^{d(\sigma,\sigma^{(0)})} W(\sigma^{(0)})~.
\l{one-gen}
\ee
Thus each sequence gets populated in one generation with a fraction which 
is same for all the sequences in a \emph{shell} of constant Hamming distance 
$d(\sigma,\sigma^{(0)})$ from the initial sequence $\sigma^{(0)}$ 
\cite{Krug:2003}. Numerical simulations of \cite{Krug:2003} showed that 
dynamical properties involving the most populated genotype 
such as the distribution of evolution times and 
number of jumps are very well described by a simplified model which ignores 
mutations for further evolution and allows the population at each sequence to 
grow with its own fitness. Thus within the shell model, the population 
${\cal Z}(\sigma,t) \sim \mu^{d(\sigma,\sigma^{(0)})} W^t (\sigma)$ 
for $t > 1$.

Here we provide an analytical understanding of the quasispecies model leading 
to the shell model. We will find that the 
expression for ${\cal Z}(\sigma,t)$ in shell model is a good 
approximation for sequences with high fitness and at short times. 
For $t > 1$, consider (\ref{linear}) for a sequence $\sigma$ in 
the shell $d$ centred about $\sigma^{(0)}$ and at a Hamming distance 
$d(\sigma,\sigma^{(0)})$ from 
the center. The sum on the right hand side of (\ref{linear}) has three 
kind of terms: (i) sequence $\sigma$ does not mutate 
({\it i.e.} $\sigma'=\sigma$ in 
the sum) so that its contribution $\propto \mu^{d(\sigma,\sigma^{(0)})}$,  
(ii) a mutation occurs in $\sigma' \in I$ where $I$ is the set of 
$\sigma'$'s lying inside shell $d$ that satisfy 
$d(\sigma,\sigma')+d(\sigma',\sigma^{(0)})=d(\sigma,\sigma^{(0)})$ resulting 
in $\mu^{d(\sigma,\sigma^{(0)})}$ dependence and 
(iii) a mutation occurs either in the sequences in the inner shells that do 
not belong to set $I$ or, in the sequences in and   
outside shell $d$ giving a term of order $\mu^{d(\sigma,\sigma^{(0)})+1}$ 
and higher. In order to obtain ${\cal Z}(\sigma,t)$ to order 
$\mu^{d(\sigma,\sigma^{(0)})}$, we can neglect the last contribution 
and iterate the population according to 
\be
{\cal Z}(\sigma,t+1)= W(\sigma){\cal Z}(\sigma,t) 
+\sum_{\sigma^{'} \in I} \mu^{d(\sigma,\sigma')} W(\sigma') {\cal Z}(\sigma',t)~.
\ee

The above equation is still coupled but one can make further simplifications 
by proceeding as follows. Let us first consider the 
zeroth shell {\it i.e.} sequence $\sigma^{(0)}$ for which we immediately have 
\be
{\cal Z}(\sigma^{(0)},t)= W^t(\sigma^{(0)})~.
\l{Zshell0}
\ee
For the sequences in the first shell for which $d(\sigma,\sigma^{(0)})=1$, 
the sequence $\sigma^{(0)}$ is the only member of set $I$ and we have
\be
{\cal Z}(\sigma,t+1)= W(\sigma){\cal Z}(\sigma,t)+\mu W^{t+1}(\sigma^{(0)})~,~d(\sigma,\sigma^{(0)})=1
\l{shell1}
\ee
Defining ${\cal Z}(\sigma,t)=W^{t}(\sigma) z(t)$, we obtain a simple 
difference equation for $z(t)$ which can be iterated to give 
\be
z(t)= \mu r+ \mu r^2 ~\frac{1- r^{t-1}}{1-r}~,~r=\frac{W(\sigma^{(0)})}{W(\sigma)}~.
\no
\ee
If $W(\sigma) > W(\sigma^{(0)})$, the population ${\cal Z}(\sigma,t)$ grows 
exponentially fast with a rate equal to its own (log) fitness 
whereas in the opposite case, this growth rate is that of the initial 
sequence. For the sequences in the first shell, one therefore obtains
\be
{\cal Z}(\sigma,t)= 
\begin{cases} 
\mu W(\sigma^{(0)}) W^{t-1}(\sigma)~,~ r < 1 \cr
\mu W^{t}(\sigma^{(0)})~,~r > 1~.
\end{cases}
\l{Zshell1}
\ee
For the succeeding shells, one can work out the population 
${\cal Z}(\sigma,t)$ 
in a manner similar to above and find that 
${\cal Z}(\sigma,t) \sim \mu^{d(\sigma,\sigma^{(0)})} 
W^t(\sigma)$ if the fitness $W(\sigma)$ 
is larger than that of all the sequences in set $I$. Such a random variable 
whose value is larger than all the ones preceding it defines a {\it record}  
\cite{Jain:2005,Krug:2005}. If the fitness $W(\sigma)$ is not a record, then 
the growth rate 
is given by the largest (log) fitness of sequences in set $I$ {\it i.e.} the 
last record value. 
 
\begin{figure}
\begin{center}
\includegraphics[angle=270,scale=0.4]{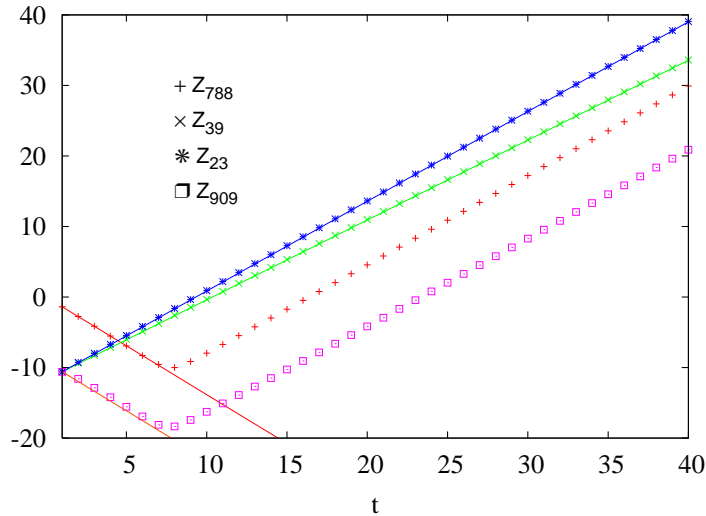}
\caption{(Color online)Comparison of the logarithmic population fraction 
$\ln {\cal Z}(\sigma,t)$ obtained by
iterating exact equation (\ref{linear}) (shown by points) and the shell model 
dynamics (\ref{Zshell0}) and (\ref{Zshell1}) (shown by lines) 
for $\mu=10^{-4}$. The 
fitness $W$ is chosen from a common exponential distribution for a 
sequence of length $L=10$. The subscript in ${\cal Z}$ gives the fitness rank 
where the largest fitness carries rank $0$ and the population initially starts 
at a sequence with rank $788$. All the other sequences are at Hamming distance 
$1$ from $\sigma^{(0)}$.}
\label{Zshell01}
\end{center}
\end{figure}

A comparison of the results obtained using above approximations 
with the exact iteration of (\ref{linear}) is shown in 
Fig.~\ref{Zshell01} for the initial sequence and some representative 
sequences in the first shell. The disagreement at large times occurs because 
(\ref{Zshell0}) and (\ref{Zshell1}) are  
obtained to leading order in $\mu$ and after some time as the population at 
the outer shells grow, the next order terms in $\mu$ start contributing. 
For the zeroth shell, this happens at time $\tau$ when the next order terms  
$\mu {\cal Z}(\sigma,\tau)$ due to sequences in shell one become as large as 
the lowest order term given by (\ref{Zshell0}) {\it i.e.} 
$W^\tau(\sigma^{(0)}) \sim \mu^2 W^{\tau}(\sigma^*)$  
where $W(\sigma^*) > W(\sigma^{(0)})$ is the largest fitness in the 
shell $1$.  
Plugging in the fitness values of relevant sequences in this expression for 
the fitness landscape in Fig.~\ref{Zshell01}, we obtain $\tau \approx 8$ 
in good agreement with the exact iteration. After time $\tau$, the population 
at $\sigma^{(0)}$ grows with the fitness $W(\sigma^*)$ until the shell $2$ 
starts contributing. This argument can be generalised 
to higher shells straightforwardly. For instance, the correction to 
(\ref{Zshell1}) is of order $\mu^3$ which arises either due to the sequences 
in the first shell but two mutations away from $\sigma$ or the sequences 
in the second shell that are one mutational distance away. The larger 
of the two contributions can be then used to estimate the time at which 
the growth rate changes. This process of slope changes of logarithmic 
population goes on until the globally fittest sequence $\sigma^{(f)}$ 
becomes most populated after which the population 
${\cal Z}(\sigma,t) \sim \mu^{d(\sigma,\sigma^{(f)})} W^t(\sigma^{(f)})$.

In Fig.~\ref{Zshell01}, the initial sequence with rank $788$ remains most 
populated until $\sim 5$ time steps after which the sequence ranked $23$ 
overtakes it and becomes the next leader. We are interested in such 
leadership changes and will use the shell model for this purpose as it 
correctly 
captures the dynamical behavior of the most populated genotype. 
Recall that in the shell model, the population at each sequence grows with 
its own intrinsic fitness for all $t >1$. This population dynamics are 
different from quasispecies in which the population of a sequence 
whose fitness is not a record grows with the fitness of the last record 
(in inner shells) at short times and the growth rate of the sequences 
change as the leading genotype changes in course of time. 
However, since the population at such sequences is always at least order $\mu$  smaller than that at the leading genotype,  the dynamics 
of the most populated sequence are not affected if the population at such sequences  is also allowed to grow with their respective fitness.

\begin{figure}
\begin{center}
\includegraphics[angle=270,scale=0.4]{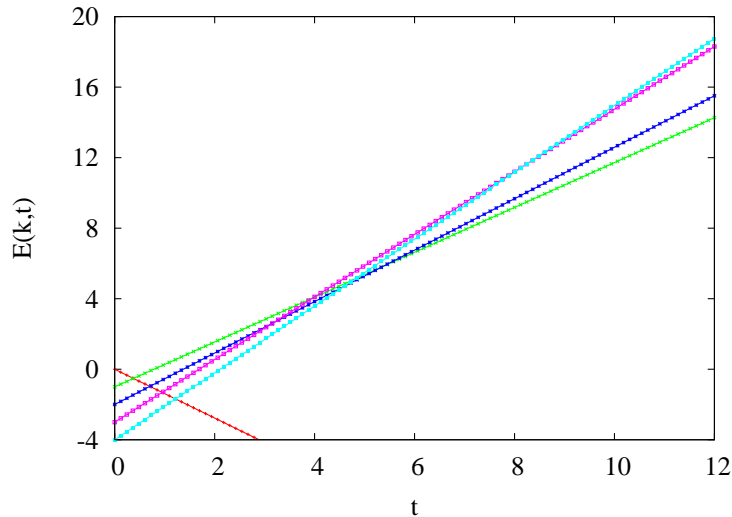}
\caption{(Color online)Shell model dynamics for population $E(k,t)=-k+F(k) t$ for 
$k=0,...,4$ for the landscape in Fig.~\ref{Zshell01}. The 
sequence with the largest fitness in shell $2$ gets bypassed by the 
corresponding sequence in shell $3$.}
\label{bypass}
\end{center}
\end{figure}

After these considerations, we arrive at the 
shell model in which the logarithmic  population obeys the following 
linear time evolution
\be
\ln {\cal Z}(\sigma,t)= \ln W(\sigma^{(0)})- 
\vert \ln \mu \vert d(\sigma,\sigma^{(0)})+ \ln W(\sigma) \; (t-1) .
\l{lines}
\ee
 Calling $F(\sigma)=\ln W(\sigma)$ and rescaling time by $|\ln \mu|$, we have 
\be
E(\sigma,t)=-d(\sigma,\sigma^{(0)}) + F(\sigma)~t
\ee
where we have absorbed the extraneous factors in the definition of 
logarithmic population $E(\sigma,t)$. Since the population 
at $t=1$ is same for sequences at constant $d(\sigma,\sigma^{(0)})=k$, 
only the sequence with the largest fitness in shell $k$ need to be 
considered \cite{Krug:2003}. If the logarithmic fitness $F$ is distributed 
according to the 
distribution $p(F)$, then the largest fitness $F(k)$ in shell $k$ which is the 
maximum of $\alpha_k={L \choose k}$ variables is distributed independently 
but non-identically with distribution 
\be
p_k(F)=\alpha_k~p(F)~\left(\int_{F_\m}^F dF'~p(F')\right)^{\alpha_k-1}
\ee
where $F_\m$ is the lower support of the distribution $p(F)$. 

\section{Statistics of jumps in shell model}
\l{jump-shell}

In the following, we will consider the shell model dynamics defined as 
\be
E(k,t)=-k+ F(k)~t
\l{ekt}
\ee
where $F(k)$ and $E(k,t)$ respectively are the fitness and population of 
the sequence with the largest fitness in shell $k=0,...,L$. 
Figure~\ref{bypass} shows the population $E(k,t)$ as a function of time 
for $k=0,...,4$ as the global maximum of the fitness 
landscape in Fig.~\ref{Zshell01} lies at a Hamming distance $4$ from the 
initial sequence. Since a line $E(k,t)$ intersects $E(k',t)$ at a time 
\be
T= \frac{k'-k}{F(k')-F(k)}~,~
\l{time}
\ee
and as $F(1) > F(0)$, the population of the sequence in shell $0$ is overtaken 
by the one in shell $1$. However to become a most populated genotype,  
it is not sufficient to have a record fitness. As Fig.~\ref{bypass} shows 
although $F(2) > F(1)$, the population $E(2,t)$ overtakes $E(1,t)$ 
later than $E(3,t)$ losing the evolutionary race. 
Thus, only those set of sequences become leader that 
manage to overtake the current leader in the least time amongst all 
contenders. Finally, the globally fittest sequence $E(4,t)$ catches up 
with $E(3,t)$ and the population localises at the global  
peak \cite{Krug:2003,Jain:2005}. 

The shell model defined by (\ref{ekt}) describes a population growing linearly 
in time with a slope equal to the fitness $F(k)$ 
chosen from the distribution $p_k(F)$. 
A simpler version of this model in which the fitness $F(k)$ are assumed to be 
independent and identically distributed (i.i.d.) variables with 
distribution $p(F)$ 
has also been considered \cite{Krug:2003,Jain:2005}. Several properties 
of this model have been recently calculated via a mapping to 
a first-passage problem \cite{Sire:2006} and considering it as a system of 
hard-core particles undergoing elastic collisions \cite{Bena:2007}. 
In particular, it has been shown 
that the average number of jumps grows as $\beta \ln L$ where the 
prefactor $\beta < 1$ and depends on the choice of $p(F)$. In both of these 
approaches, the initial condition (the intercept) of $E(k,t)$ 
is not fixed and is a uniformly distributed random number on the real line. 
In this article, we present a way to calculate average number ${\cal J}$ of 
changes in $k^*$ which 
respects the discreteness of the underlying genotypic space. We will 
perform the calculations for the shell model for which 
the fitness is non-identically distributed, although our 
method is readily applicable to the i.i.d. model also. However, we mention 
that the prefactor $\beta$ calculated using our method turns out to be the 
same as in the analysis of \cite{Sire:2006,Bena:2007}. 

\subsection{General formulae}

To calculate the statistics of jumps, we need two basic 
distributions: (i) the probability $P_k(F,t)$ that the most 
populated sequence in shell 
$k$ with fitness $F$ is the leader at time $t$ and (ii) the probability 
$W_{k',k}(F,t) dt$ with which 
this sequence is overtaken by the most populated sequence in shell $k' (> k)$ 
between time $t$ and $t+dt$. 
The distribution $P_k(F,t)$ requires that the population $E(j,t) < E(k,t)$, 
$j \neq k$ which implies that the fitness $F(j) < F+ (j-k)/t$. 
Since the fitnesses of the most populated sequence in each shell are 
independent random variables, we have
\be
P_{k}(F,t)=p_k(F)~ \prod_{\substack{j=0\\j \neq k}}^L \int_{F_\m}^{F+\frac{j-k}{t}} dF'~p_j(F')~. 
\l{max}
\ee
To find $W_{k',k}(F,t)$, we need to determine the fitness of the most 
populated sequence in shell $k' (> k)$ which can contribute to the overtaking 
event. Let us denote the location of the leader at time $t$ by $E(t)$, 
\be
E(t)=-k+F~t ~.
\no
\ee
Then the sequence in shell $k'$ can overtake the $k$th one at $t$ if the 
fitness $F(k')=(E(t)+k')/t$. Similarly, the sequence in the $k$th shell can 
be overtaken at $t+dt$, $dt/t \to 0$ if 
\be
F(k')=\frac{E(t+dt)+k'}{t+dt}= F+\frac{k'-k}{t}-\frac{k'-k}{t^2} dt+{\cal O}(dt^2)~. \no
\ee
Thus the probability that a sequence in shell $k$ is overtaken by a sequence 
in shell $k'$ between time $t$ and $t+dt$ is given by
\be
W_{k',k}(F,t) dt=\int_ {F+\frac{k'-k}{t}-\frac{k'-k}{t^2} dt}^{F+\frac{k'-k}{t}} 
dF~p_k(F) \approx \frac{k'-k}{t^2} ~p_{k'} \left(F+ \frac{k'-k}{t} \right)~dt~,~k' > k~.
\l{coll}
\ee
Finally using the distributions defined 
in (\ref{max}) and (\ref{coll}), we can write 
the probability ${\cal P}_{k',k}(t)$ that the 
most populated sequence in shell $k'$ overtakes the one in shell $k$ at time 
$t$ as
\be
{\cal P}_{k',k}(t)= \int_{F_\m}^{F_\M} dF~W_{k',k}(F,t)~P_{k}(F,t)
\ee
where $F_\M$ is the upper support of the distribution $p(F)$. 

Depending on the quantity of interest, one can either integrate over time 
or sum over a space variable in ${\cal P}_{k',k}(t)$. Often the 
experimental data such as a morphological feature \cite{Benton:2001} or 
average fitness \cite{Elena:2003a} are plotted as a function of 
time and can be used to find the number of jumps in time. Therefore it 
is useful to consider the distribution ${\cal J}(t)$ of a jump to occur 
at time $t$ which can be found by summing ${\cal P}_{k',k}(t)$ 
over $k$ and $k'$, 
\be
{\cal J}(t)=\sum_{k=0}^{L} \sum_{k'=k}^{L}{\cal P}_{k',k}(t)~.
\l{Jt}
\ee
The relationship between the overtaken and the overtaking sequence can 
be deduced by computing the 
jump distribution ${\cal J}_k$ that the most populated sequence 
in shell $k$ is a jump given by
\be
{\cal J}_k=\sum_{k'=k}^{L} \int_0^\infty dt~{\cal P}_{k',k}(t)~.
\l{Jk}
\ee
The average number ${\cal J}$ of jumps can be obtained by either integrating 
${\cal J}(t)$ over time or summing ${\cal J}_k$ over $k$. 

In the previous works on shell model 
\cite{Krug:2003,Jain:2005,Krug:2005}, several properties 
of the jumps have been  studied numerically when the fitness $F$ is 
distributed according to an exponential or Gaussian distribution. In the 
following subsections, we will use the expressions derived above 
for the exponential case which lends itself to a detailed analysis and then 
give some results for fitness 
distributions decaying faster than an exponential. 

\subsection{Exponentially distributed fitness}

We consider $p(F)=e^{-F}$ for which 
the largest fitness in shell $k$ is distributed according to 
\be
p_k(F)=\alpha_k ~e^{-F} (1-e^{-F})^{\alpha_k-1}~.
\l{pkF}
\ee
Using this in (\ref{max}), we obtain the distribution that the leader with 
fitness $F$ is in shell $k$, 
\bea
P_{k}(F,t) &=& \alpha_k \left(\frac{e^{-F}}{1-e^{-F}} \right) \prod_{j=0}^L 
(1-e^{-F+\frac{k-j}{t}})^{\alpha_j} \no \\
&\approx&\alpha_k ~e^{-F}~ e^{-(1+e^{-\frac{1}{t}})^L ~e^{-F+\frac{k}{t}}}
\eea
where the last expression is obtained by exponentiating the product and 
keeping only the leading order terms in the expansion. Similarly, the 
overtaking rate (\ref{coll}) can be written as 
\bea
W_{k',k}(F,t) &=&\alpha_{k'} \left(\frac{k'-k}{t^2}\right) e^{-F+\frac{k-k'}{t}} ~(1-e^{-F+\frac{k-k'}{t}})^{\alpha_{k'}-1}  \no \\
&\approx&\alpha_{k'} \left(\frac{k'-k}{t^2}\right)e^{-F+\frac{k-k'}{t}}~e^{-\alpha_{k'}e^{-F+\frac{k-k'}{t}}}~.
\eea
Then the probability that the population in the $k'$th shell exceeds the 
population in the $k$th shell at time $t$ is given by 
\be
{\cal P}_{k',k}(t)= \alpha_k~ \alpha_{k'}~ e^{\frac{k-k'}{t}}\left(\frac{k'-k}{t^2}\right) \int_0^1 dz ~z ~e^{-z ~[\alpha_{k'}e^{-\frac{k'}{t}}+ (1+e^{-\frac{1}{t}})^L ] e^{\frac{k}{t}}}~.
\ee
Neglecting the first term in the exponent of the exponential in the integrand 
and carrying out the integral for large $L$, we finally obtain 
\be
{\cal P}_{k',k}(t) \approx \frac{\alpha_{k'}~e^{-\frac{k'}{t}}}{(1+e^{-\frac{1}{t}})^L}~
\frac{\alpha_{k}~e^{-\frac{k}{t}}}{(1+e^{-\frac{1}{t}})^L}~\left(\frac{k'-k}{t^2}\right)~.
\l{basic} 
\ee
\begin{figure}
\begin{center}
\includegraphics[angle=270,scale=0.35]{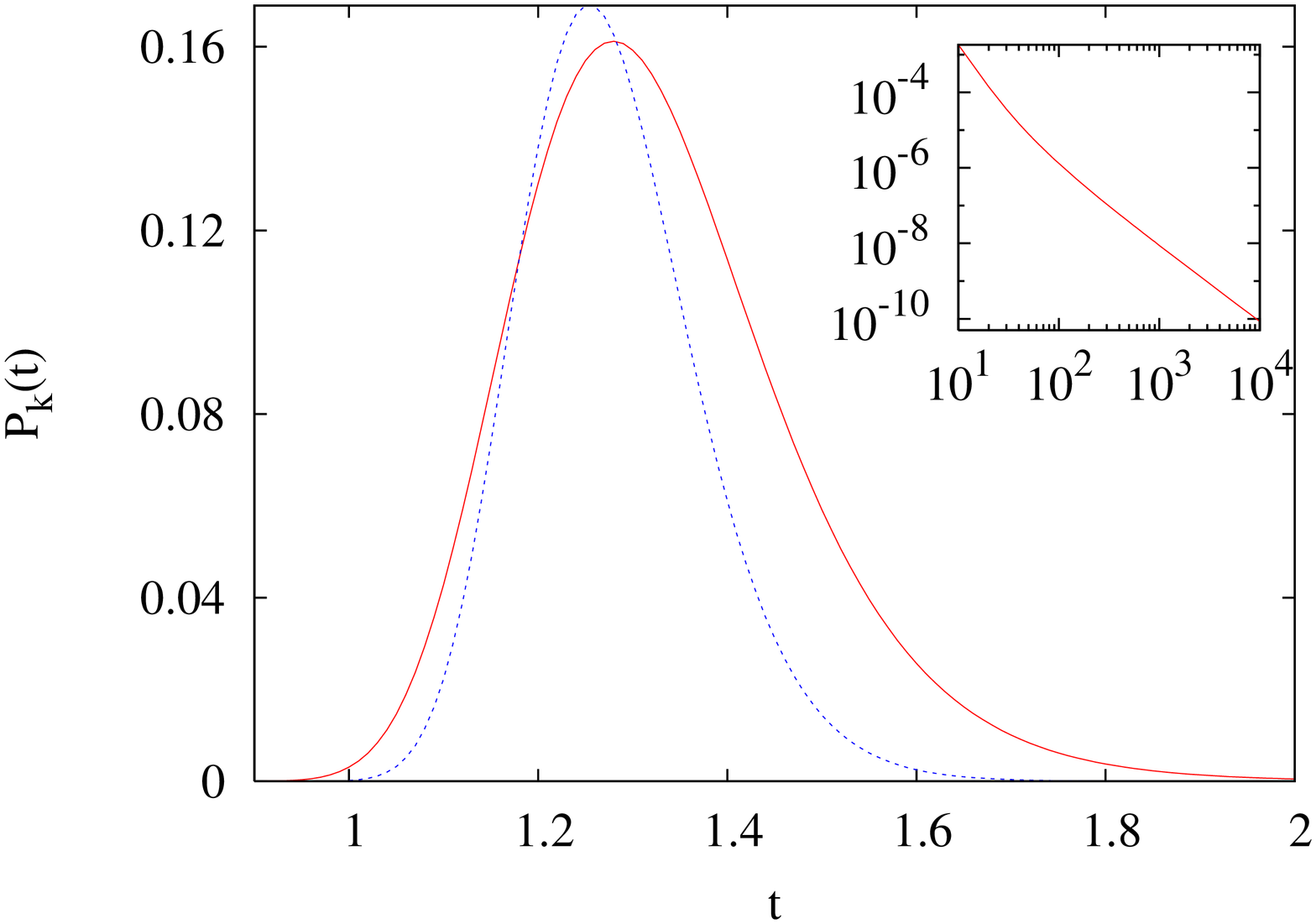}
\includegraphics[angle=270,scale=0.35]{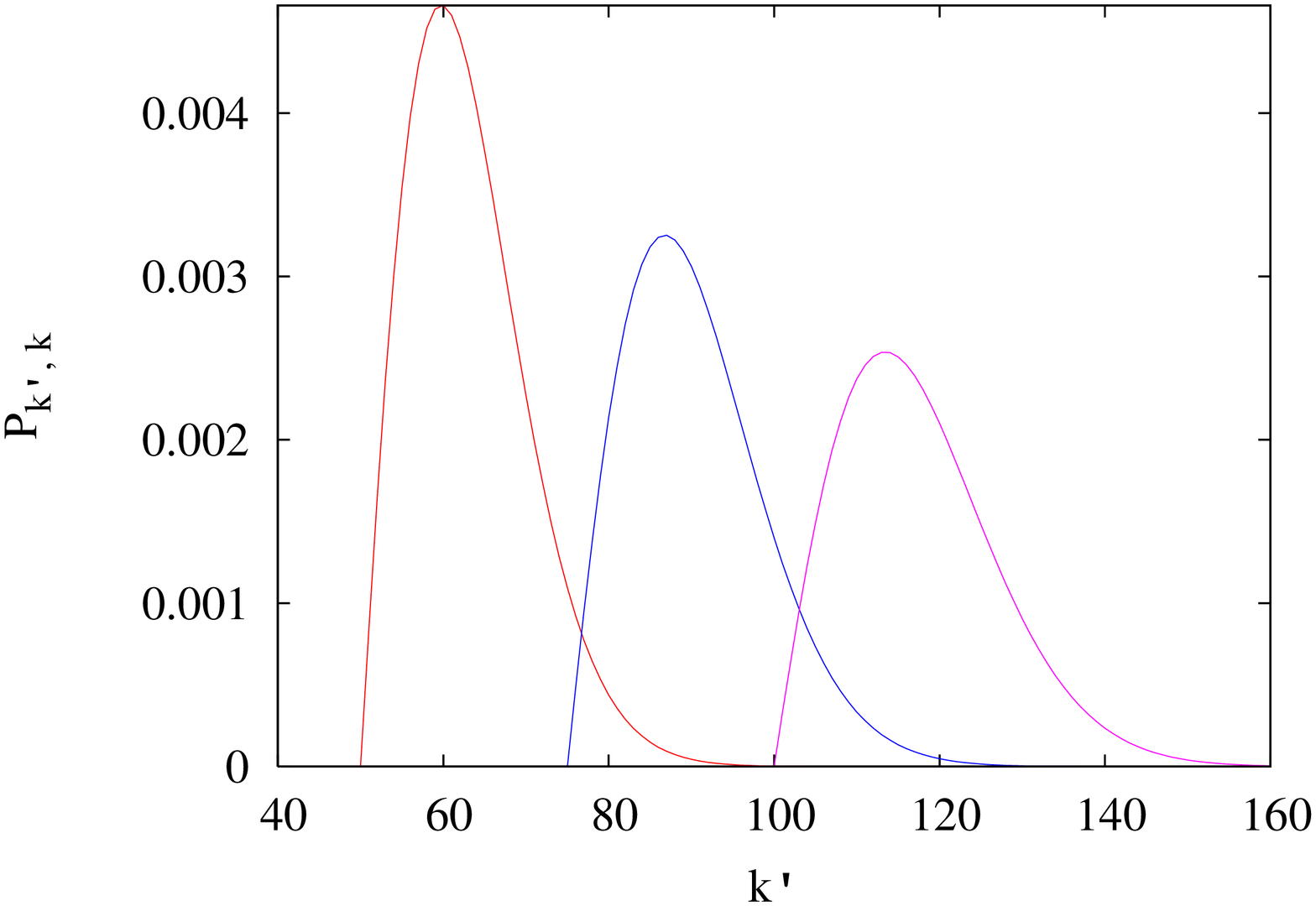}
\caption{(Color online)Jump statistics for $p(F)=e^{-F}$. 
Left: Distribution ${\cal P}_k(t)$ given by (\ref{Pkt}) 
for fixed $k/L=0.3$ in support 
of (\ref{Tmp}) for $L=500$ (solid) and $1000$ (dotted). The inset shows the 
$1/t^2$ dependence of ${\cal P}_k(t)$ for $k=450, L=1000$. 
Right: Distribution ${\cal P}_{k',k}$ given by (\ref{Pkpk}) for $k=50, 75$ and 
$100$ (left to right) and $L=1000$.}
\label{expo}
\end{center}
\end{figure}
We will now use this expression to calculate jump statistics. 

\noindent{\it Temporal behavior-.} 
Let us first consider the distribution ${\cal P}_{k}(t)$ of a jump to 
occur in the $k$th shell at time $t$.
Summing ${\cal P}_{k',k}(t)$ over $k'$, we obtain \cite{Gradshteyn:1980}
\be
{\cal P}_{k}(t)=\frac{\alpha_k~e^{-\frac{k}{t}}}{t^2(1+e^{-\frac{1}{t}})^{2L}}\sum_{k'=k}^L (k'-k) ~\alpha_{k'} e^{-\frac{k'}{t}} =\frac{\alpha_k~e^{-\frac{k}{t}}}{t^2(1+e^{-\frac{1}{t}})^{2L}} \frac{\Gamma(L+1) e^{-\frac{k+1}{t}}}{\Gamma(L-k) \Gamma(k+2)} ~{_2F_1}(2,k+1-L;k+2;-e^{-\frac{1}{t}})
\l{Pkt}
\ee
where $\Gamma(n)$ is the gamma function and ${_2F_1}(a,b;c;z)$ is the 
hypergeometric function. We point out that the distribution 
${\cal P}_{k}(t)$ gives the probability that $E(k,t)$ is {\it overtaken} at 
$t$ and hence differs from the distribution that 
$E(k,t)$ is the largest at time $t$ considered in \cite{Jain:2005}. 
The function ${\cal P}_{k}(t)$ is plotted as a function of 
time for various values of $k$ in Fig.~\ref{expo}. 
To gain some insight into the 
behavior of this distribution, we calculate the above sum using saddle point 
approximation. Using the Stirling's formula for binomial coefficient
\be
{L \choose k} \approx \sqrt{\frac{L}{2 \pi k (L-k)}} ~\frac{L^L}{k^k (L-k)^{L-k}}
\l{stirling}
\ee
in (\ref{Pkt}) for $\alpha_{k'}$ and approximating the sum over $k'$ by an 
integral, we have 
\be
\int_k^L dk'~(k'-k)~e^{-f(k')} \approx \frac{\sqrt{\pi} e^{-\frac{k'_0}{t}}}{f''(k'_0)} {L \choose k'_0} \left[ \Delta(k) [{\rm erf}(\Delta(k))-{\rm erf}(\Delta(L))]+\sqrt{\frac{1}{\pi}}~ [e^{-\Delta^2(k)}-e^{-\Delta^2(L)}] \right]
\l{Aint} \no
\ee
where we have estimated the integral using the saddle point method. In the 
above expression, $f(k')=(k'/t)-\ln \alpha_{k'}$,  
$f''(k'_0)$ is the second derivative of $f(k')$ evaluated at the minimum 
$k'_0$ of the function $f(k')$ and the deviation $\Delta(k')=(k'-k'_0) 
\sqrt{f''(k'_0)/2}$. Explicitly, 
\be
k'_0=\frac{L}{1+e^{\frac{1}{t}}} ~,~ f''(k'_0)=\frac{L}{k'_0 (L-k'_0)}=
\frac{(1+e^{\frac{1}{t}})(1+e^{\frac{-1}{t}})}{L}~,~\Delta(k)=(k-k'_0) 
\sqrt{\frac{f''(k'_0)}{2}}~. \no
\ee
Approximating the factor $\alpha_k e^{-\frac{k}{t}}$ in (\ref{Pkt}) also by 
a Gaussian in a manner similar to above, we finally have 
\be
{\cal P}_k(t)= \frac{e^{-\Delta^2(k)}}{2 \sqrt{\pi} t^2}~\left[ \Delta(k) [{\rm erf}(\Delta(k))-{\rm erf}(\Delta(L))]+\sqrt{\frac{1}{\pi}}~ [e^{-\Delta^2(k)}-e^{-\Delta^2(L)}] \right]~.
\l{Pkta}
\ee

Using this expression, it is easy to obtain the typical shell location of the 
sequence overtaken at time $t$ by integrating $k {\cal P}_k(t)$ over $k$. 
Since ${\cal P}_k(t)$ is normalised to ${\cal J}(t)$, we find that given an 
overtaking event occurs at $t$, the average 
location of the overtaken sequence scales as $k_0'$ with a standard deviation 
of order $1/\sqrt{f''(k'_0)}$ about it. This length scale can also be 
expressed in terms 
of time for fixed $k$ as the distribution ${\cal P}_k(t)$ is maximised 
at $\Delta(k)=0$ or $k'_0(t)=k$. 
Thus for large $L$, the time $T_{\rm{m.p.}}$ at which ${\cal P}_k(t)$ is {\it most probable} is given by 
\be
T_{\rm{m.p.}}=\left[\ln \left(\frac{L-k}{k} \right) \right]^{-1}~.
\l{Tmp}
\ee
This means that a sequence with $k \sim {\cal O}(L)$ is most likely to be 
overtaken in a time of order unity. This fact is also expressed in 
Fig.~\ref{expo} which shows the distribution ${\cal P}_k(t)$ at fixed 
$k/L$ as a function of time. The time scale $T_{\rm{m.p.}}$ can be 
understood by a simple argument which estimates 
the intersection time given by (\ref{lines}). This argument is analogous to 
that used in \cite{Krug:2003,Jain:2005,Jain:2007a} where the fitness 
difference in the denominator of (\ref{lines}) is given by the 
typical value of the fitness gap which probes the rare events.  As we are 
interested in the 
most likely events, the denominator is approximated by the difference in the 
average value of the largest fitness in shell $k'$ and $k$. 
From (\ref{pkF}), we see that the average largest fitness goes as 
$\ln \alpha_k$. Since typically the most populated 
sequence in shell $k$ is overtaken by a sequence located ${\cal O}(\sqrt{k})$ 
distance away \cite{Jain:2005} (also see below), the 
numerator of (\ref{lines}) scales as $\sqrt{k}$ and the 
denominator $\ln (\alpha_k'/\alpha_k )$ on using the Stirling's formula 
turns out to be $\sqrt{k} \ln((L-k)/k)$ thus leading to (\ref{Tmp}).

Although the most probable value of the overtaking time is of order one, the 
{\it average} overtaking time is infinite due to the fat tail of the 
distribution ${\cal P}_k(t)$. For $t \gg 1$, we can approximate $\Delta(k)$ by 
$(2 k-L)/{\sqrt{2 L}}$ and using the 
asymptotic expansion of error function for large argument \cite{Arfken:1985}, 
\be
{\rm erf}(x)= \frac{2}{\sqrt{\pi}} \int_0^x dy~e^{-y^2}=1-\frac{e^{-x^2}}{\sqrt{\pi} x} \left(1- \frac{1}{2 x^2}+...\right)
\l{erfx}
\ee
we obtain
\be
{\cal P}_k(t) \approx \sqrt{\frac{L}{2 \pi}}~\frac{\epsilon}{t^2}~e^{-L \epsilon^2/2}~,~\epsilon=1-\frac{2k}{L} > 0~.
\l{larget}
\ee
Due to the $t^{-2}$ behavior at large times 
shown in the inset of Fig.~\ref{expo}, the mean time diverges for 
any $L$ and $k < L/2$ (see below).  
The tail of this distribution is exponentially suppressed in $L$ for 
finite $\epsilon$ but goes as $\sqrt{L} \epsilon/t^2$ for $k$ close to $L/2$. 
The late time behavior above is also obtainable from (\ref{lines}) by a 
simple change of variables \cite{Krug:2003,Jain:2005}.

After performing the integral over $k$ in (\ref{Pkta}), we obtain 
\be 
{\cal J}(t)=\sqrt{\frac{L}{4 \pi}}~\frac{1}{t^2} ~\mathrm{sech} \left(\frac{1}{2t} \right)~.
\l{Jt-shell}
\ee
Integrating ${\cal J}(t)$ 
over time from $0$ to infinity, we find that the average number of jumps 
grows as $\sqrt{L \pi}/2$. 

\noindent{\it Spatial behavior-.} 
One can also find the probability ${\cal J}(k)$ that the most populated 
sequence in the $k$th shell is a jump. As we are not interested in temporal 
distribution, the integral over time in (\ref{basic}) can be carried out 
to give the probability that the sequence in 
shell $k$ is overtaken by that in shell $k'$, 
\be
{\cal P}_{k',k}=\int_0^{\infty} dt~{\cal P}_{k',k}(t) =
\frac{k'-k}{k'+k} ~{L \choose k} {L \choose k'} {_2F_1}(k+k',2L;k+k'+1;-1)~. 
\l{Pkpk}
\ee
This distribution is shown for some representative parameters in 
Fig.~\ref{expo}. Approximating the integrand ${\cal P}_{k',k}(t)$ 
in the above equation by a 
Gaussian centred about inverse time $t^{-1}=-\ln((k'+k)/(2L-k'-k))$ and 
carrying out the integral, we obtain
\be
{\cal P}_{k',k} \approx
\frac{L (k'-k)}{(k+k') (2L-k'-k)} {L \choose k} {L \choose k'} {2L \choose k'+k}^{-1} 
\left[1+{\rm erf} \left(\sqrt{\frac{(k+k') (2L-k-k')}{4L}} \ln \left(\frac{2L-k'-k}{k'+k} \right)  \right)  \right] ~. 
\label{erf-ln}
\ee
The argument of the error function changes sign when $k'+k=L$. For $k'+k > L$, 
 the argument is negative and of order $\sqrt{L} \gg 1$. Using (\ref{erfx}),   
we find that the last factor in (\ref{erf-ln}) is exponentially small in $L$ 
for $k' > L/2, k < k'$.  
Thus the probability that the overtaking sequence $k'$ lies beyond 
the shell $L/2$ is negligible. This is understandable as the globally fittest 
sequence is typically located in the shell $k=L/2$ \cite{Jain:2005}. 
For $k,k' < L/2$, the error function in (\ref{erf-ln}) can be approximated 
by unity, and the probability distribution ${\cal P}_{k',k}$ can be 
further simplified to give
\be
{\cal P}_{k',k} \approx \sqrt{\frac{L}{\pi k (L-k)}}~\left(\frac{k'-k}{2 k}\right)~e^{-\frac{L (k'-k)^2}{4k (L-k)}}~,~k< k' < L/2
\ee
where we have used the Gaussian approximation for the binomial coefficients. 
This form of the distribution implies that the overtaking sequence $k'$ is 
located within ${\cal O}(\sqrt{k})$ distance of the overtaken sequence $k$. 
Thus the typical spacing between successive jumps 
for large $k$ is roughly constant and goes as $\sqrt{L}$ as seen in the 
numerical simulations of \cite{Jain:2005}. 
The jump distribution ${\cal J}_k$ for a jump to occur in shell $k$ 
is obtained by integrating over $k'$ and we have
\be
{\cal J}_k \approx \sqrt{\frac{L}{\pi k (L-k)}} ~\theta_H \left(\frac{L}{2}-k \right)
\ee
where $\theta_H$ is the Heaviside step function. Thus the distribution 
${\cal J}_k$ decays as $k^{-1/2}$ for $k \ll L$ in accordance with the 
numerical results of \cite{Jain:2005,Krug:2005}. Integrating the preceding 
equation over $k$, we find 
that the average number of jumps are given as $\sqrt{L \pi}/2$ 
in agreement with the previous calculation. 

\subsection{Gumbel-distributed shell fitness}

We now consider $p(F)=A e^{-F^{\gamma}}$, $\gamma >0, A=\gamma/\Gamma(1/\gamma)$ for which the 
distribution $p_k(F)$ of the largest of $\alpha_k$ random variables for large 
$L$ has the Gumbel distribution as the limiting form \cite{David:1970},
\be
p_k(F) = \frac{1}{C_k}~e^{-\frac{F-B_k}{C_k}}~e^{-e^{-\frac{F-B_k}{C_k}}}
\l{scal}
\ee
where 
\be
B_k= \left[\ln \left(\frac{\alpha_k}{\Gamma(1/\gamma)} \right) \right]^{1/\gamma} ~,~
C_k= \frac{1}{\gamma} 
\left[ \ln \left(\frac{\alpha_k}{\Gamma(1/\gamma)} \right)\right]^{(1-\gamma)/\gamma}~.
\ee
We will show that the tail of the distribution ${\cal P}_k(t)$ 
decays as $1/t^2$ and the average number 
of jumps scales as $\sqrt{L}$ for any $\gamma > 0$. 

The large time behavior of ${\cal P}_k(t)$ can be found by 
taking $t \to \infty$ limit in (\ref{max}) and (\ref{coll}) except for the 
$1/t^2$ factor in rate $W_{k',k}$. We thus have 
\be
{\cal P}_k(t) \approx \sum_{k'=k}^L \frac{k'-k}{t^2} ~
\frac{e^{\frac{B_k'}{C_k'}+\frac{B_k}{C_k}}}{C_k' C_k}~ 
\int_0^\infty dF~e^{-\frac{F}{C_k}}  e^{-\frac{F}{C_k'}}  
e^{-\sum_{j=0}^L e^{\frac{B_j}{C_j}} e^{-\frac{F}{C_j}}}  ~,~t \gg T_{\rm{m.p.}}\l{gumbel}
\ee
after using the approximations similar to those used in arriving at 
(\ref{basic}). The sum under the integral sign can be computed by saddle 
point approximation so that the integral (up to scale factors) is writeable as 
\be
\int_0^{\infty} dz ~
z^{\left(\frac{\ln \alpha_k'}{L \ln 2} \right)^{\frac{\gamma-1}{\gamma}}+\left(\frac{\ln \alpha_k}{L \ln 2} \right)^{\frac{\gamma-1}{\gamma}} -1} e^{-z} \no
\ee
where we have neglected the logarithm of $\Gamma(1/\gamma)$ in $B_k$ and 
$C_k$ for large $L$. Since we are interested in large times when $k \to L/2$, 
to leading order in $\epsilon=1-(2k/L)$, we find $\ln \alpha_k/L \ln 2 
\approx 1$ thus simplifying the expression for ${\cal P}_k(t)$ to yield
\bea
{\cal P}_k(t) &=& \left(\frac{\gamma}{t} \right)^2 
\left(\frac{\pi L}{2} \right)^{\gamma-1} (L \ln 2)^{\frac{\gamma-1}{\gamma}} 
\left(\frac{\alpha_k}{2^L}\right)^{\gamma} \sum_{k'=k}^L (k'-k) \left(\frac{\alpha_k'}{2^L}\right)^{\gamma} \no\\
&=& \sqrt{\frac{L}{2 \pi \gamma}} ~\frac{\gamma^2 \epsilon}{t^2} ~
(L \ln 2)^{\frac{\gamma-1}{\gamma}}~,~t \gg T_{\rm{m.p.}}
\eea
where the last expression is obtained by estimating the sum over $k'$ 
using Gaussian approximation. For small $\epsilon$ ($\sim 1/\sqrt{L}$)  
 and $\gamma=1$, 
this expression reduces to (\ref{larget}) for exponential distribution. 
For $\gamma \neq 1$, the 
$L$ dependence in the preceding equation consists of two factors, the 
first one arising because the typical 
separation $ k'-k \sim {\cal O}(\sqrt{L})$ for large $k$ 
\cite{Krug:2003,Jain:2005} and the last factor is due to 
$1/C_k$, $k=L/2$ left after scaling the fitness $F$ in (\ref{gumbel}) 
 by $C_{k'}$. 

In order to find the average number of jumps, we first need to calculate  
$T_{\rm{m.p.}}$ for arbitrary $\gamma$. As argued in the last subsection, 
$T_{\rm{m.p.}}^{-1}$ 
is given as the derivative (with respect to $k$) of the average 
shell fitness. From the scaling form (\ref{scal}), we see that the 
average shell fitness is proportional to $B_k$. For $k$ of order $L$, 
its derivative grows as $C_k$, $k \sim L$. Then integrating ${\cal P}_k(t)$ 
over $k$ and $t$, we find that the average number of jumps 
scale as $\sqrt{L}$ for all $\gamma > 0$. This result is also consistent 
with the simulations for the Gaussian distribution in \cite{Krug:2005}.

\section{Jump distribution for finite populations}
\l{jump-finite}

We will now consider the dynamics of a population of 
$N$ individuals evolving according to the Wright-Fisher dynamics described in 
Section \ref{models}. Unlike the infinite population, a 
population of size $N$ initially localised at 
sequence $\sigma^{(0)}$ can spread up to a finite 
distance. This is since the typical fraction of the population at a sequence 
$\sigma$ in one generation is given as 
$\mu^{d(\sigma,\sigma^{(0)})}$ but as this fraction is bounded below by $1/N$, 
it follows that the mutational spread $d_{\rm{eff}}=\ln N/|\ln \mu|$ 
for a finite population. Thus while all the mutants are available in one 
generation for quasispecies (see (\ref{one-gen})), only a finite number is 
present at any time in real populations. 
 If a fitter sequence is available within this effective distance, the 
population behaves like a 
quasispecies and evolves deterministically. This is possible for 
$d_{\rm{eff}} \geq 1$ and at short times \cite{Jain:2007a}. However 
at long times, any finite population can 
get trapped at a local peak if a fitter sequence lies farther out than 
$d_{\rm{eff}}$. In such an event, the population escapes the local peak 
via the process of 
stochastic tunneling which takes a time given by 
\cite{Iwasa:2004,Weinreich:2005}
\be
\Delta T= \left[N \mu^2 L \left(\frac{W_{\rm{loc,f}}-W_{\rm{loc,i}}}{W_{\rm{loc,f}} ~W_{\rm{loc,i}}} \right)\right]^{-1}
\ee
for $d_{\rm{eff}}=1$ where $W_{\rm{loc,[i,f]}}$ refers to the fitness of the 
initial and final local peak separated by two mutations. During this time, 
most of the population stays at the local peak with fitness $W_{\rm{loc,i}}$ 
but a few less-fit mutants are produced by single mutations. When some of 
these mutants further acquire a favorable mutation, then the whole 
population quickly jumps to the next local peak with higher fitness 
$W_{\rm{loc,f}}$. 
The physical process involved when a jump occurs in a finite population 
is thus different from that in the quasispecies case. In the latter case, 
each local peak is already populated albeit with a small frequency and a 
jump occurs when the population at a fitter sequence overtakes the current 
leader.

\begin{figure}
\begin{center}
\includegraphics[angle=270,scale=0.4]{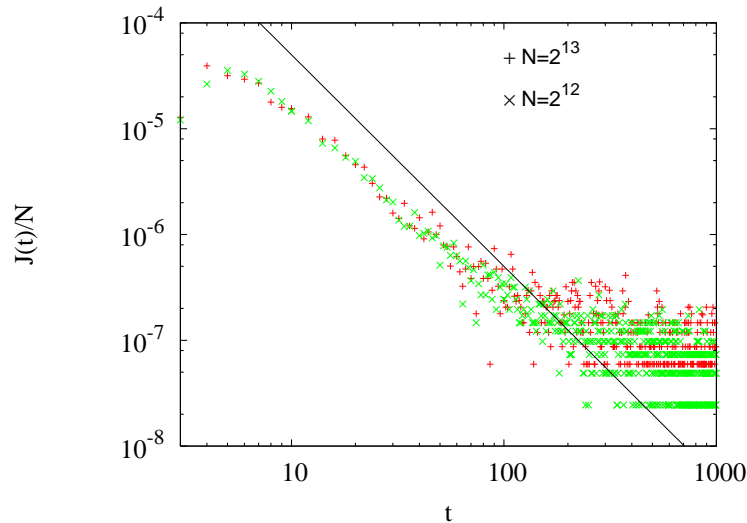}
\caption{(Color online)Scaled jump distribution $J(t)/N$ for a finite population with 
$N$ individuals to show the $1/t^2$ dependence. 
Here $L=6, \mu=0.0004, p(W)=W^{-2}$ and 
the data have been averaged over $500$ fitness landscapes with $10$ histories 
each.}
\label{finjump}
\end{center}
\end{figure}

We are interested in the jump distribution of large 
finite populations with $d_{\rm{eff}}$ close to one. 
At large times when the tunneling 
drives the dynamics, we expect the density $J(t)$ of jumps at $t$ to 
scale as 
\be
J(t) \sim \frac{1}{\Delta T} \sim 
N \mu^2 L \frac{\Delta w (t)}{w_i^2(t)}
\ee
where $w_i(t)$ is the typical fitness of the local peak visited at $t$ 
separated by a better peak with fitness difference $\Delta w (t)$. We 
expect $\Delta w$ to decrease and $w_i$ to increase with time as 
higher peaks are explored. However in the absence of an argument for these 
time dependences, we present our preliminary numerical results here.  
In the shell model, one does not have to deal with the whole genotypic 
space consisting of $2^L$ sites and it suffices to work with the 
$L$ shells thus reducing the computational effort enormously \cite{Krug:2003}. 
However such a rotational symmetry is not present 
 for the finite population problem so we 
are able to handle only small values of $L$.
Our numerical 
results for large finite populations and small $L$ 
are shown in Fig.~\ref{finjump} 
for exponentially distributed log fitness $F$ or $p(W)=W^{-2}$ 
as in the last sections. 
The data in Fig.~\ref{finjump} are averaged over several 
histories as the evolutionary trajectories are not deterministic for 
finite populations \cite{Jain:2007a}. For fixed $\mu$ and $L$, we find that 
at long times, 
\be 
J(t) \sim \frac{N}{t^2}
\l{Jt-fin}
\ee
which decays the same way as in the quasispecies model, 
\be
{\cal J}(t)\sim \left(\frac{\ln \mu}{t} \right)^2 \sqrt{L}~,~t \gg 1 
\ee
where we have reinstated the $\mu$ dependence.

\section{Conclusions}

In this article, we discussed the evolution of 
asexual population on rugged fitness 
landscapes with many local optima separated by valleys. 
We focused on 
the statistical properties of the most populated genotype which changes 
as the population locates better peaks in the fitness landscape. 
These properties were calculated exactly within a shell model which 
was derived systematically from the Eigen's quasispecies model for 
infinite populations. We showed that the expression for the population 
frequency within shell model approximates the quasispecies solution well 
for highly fit sequences and at short times only. However, the two 
models are equivalent as regards the statistics of the most populated 
genotype. 
We computed the average number of jumps in the shell model and found that it 
grows as $\sqrt{L}$, $L$ being the sequence 
length, for fitness distributions decaying as exponential or faster. 
The jump distribution in time was shown to decay as $t^{-2}$. 
This dependence is also seen numerically 
for the finite population but a satisfactory explanation for this 
case is presently missing. A more detailed analysis of the finite population 
properties will be taken up in the future.

Acknowledgement: The author thanks J. Krug for useful comments, Israel 
Science Foundation for financial support and KITP (Santa Barbara) for 
hospitality where this work was initiated. 


\end{document}